\documentclass[a4paper,11pt]{article}
\pdfoutput=1 

\usepackage{jheppub} 

\usepackage[T1]{fontenc} 
\newcommand{\chushi}[1]{}

\usepackage{amsfonts}
\newcommand{\drawsquare}[2]{\hbox{%
\rule{#2pt}{#1pt}\hskip-#2pt
\rule{#1pt}{#2pt}\hskip-#1pt
\rule[#1pt]{#1pt}{#2pt}}\rule[#1pt]{#2pt}{#2pt}\hskip-#2pt
\rule{#2pt}{#1pt}}

\newcommand{\Yfund}{\raisebox{-.5pt}{\drawsquare{6.5}{0.4}}}
\newcommand{\Ysymm}{\Yfund\hskip-0.4pt%
                    \Yfund}
\def\symm{\Ysymm}

\def\drawbox#1#2{\hrule height#2pt
        \hbox{\vrule width#2pt height#1pt \kern#1pt
              \vrule width#2pt}
              \hrule height#2pt}

\def\Fund#1#2{\vcenter{\vbox{\drawbox{#1}{#2}}}}
\def\Asym#1#2{\vcenter{\vbox{\drawbox{#1}{#2}
              \kern-#2pt       
              \drawbox{#1}{#2}}}}

\def\fund{\Fund{6.4}{0.3}}

\title{\boldmath Very light dilaton and naturally light Higgs boson}

\author{Deog Ki Hong}
\affiliation{Department of Physics, Pusan National University,
             Busan 46241, Korea}
   
\emailAdd{dkhong@pusan.ac.kr}

\abstract{
We study very light dilaton, arising from a scale-invariant ultraviolet theory of the Higgs sector in the standard model  of particle physics. 
Imposing the scale symmetry below the ultraviolet scale of the Higgs sector, we alleviate the fine-tuning problem associated with the Higgs mass. When the electroweak symmetry is spontaneously broken radiatively  {\it {\`a} la} Coleman-Weinberg, the dilaton develops a vacuum expectation value away from the origin to give an extra contribution  to the Higgs potential so that the Higgs mass becomes naturally around the electroweak scale. The ultraviolet scale of the Higgs sector can be therefore much higher than the electroweak scale, as the dilaton drives the Higgs mass to the electroweak scale.  We also show that the light dilaton in this scenario is a good candidate for dark matter of mass $m_D\sim 1~{\rm eV}-10~{\rm keV}$, if the ultraviolet scale is about $10-100~{\rm TeV}$. Finally we propose a dilaton-assisted composite Higgs model to realize our scenario. In addition to the light dilaton the model predicts a heavy ${\rm U}(1)$ axial vector boson and two massive, oppositely charged, pseudo Nambu-Goldstone bosons, which might be accessible at LHC.}


\arxivnumber{1703.05081}

\preprint{\hbox{\tt PNUTP-17/A01} }

\begin{document}

\maketitle
\flushbottom

\section{Introduction}
\label{sec;intro}
The standard model (SM) of particle physics, which has been very successful in describing the interactions of elementary particles, is finally completed by the discovery of its last missing piece, the Higgs particle, at the large hadron collider (LHC)~\cite{Aad:2012tfa,Chatrchyan:2012xdj}. The properties of the Higgs particle are measured to be consistent with the standard model prediction, better than at the percent level 
by the subsequent experiments~\cite{CMS:2017rli,ATLAS:2017myr}. 
But, nonetheless, the SM is widely regarded as  an effective theory below the electroweak scale $\sim1~{\rm TeV}$, set by the vacuum expectation value (vev) of Higgs fields.  Since 
the SM does not have any obvious symmetry to protect the mass of Higgs particle, which is very sensitive to short distance physics, it needs to be highly fine-tuned, if the ultraviolet (UV) scale of Higgs physics is much higher than TeV~\cite{'tHooft:1979bh}.
New physics at TeV is hence currently actively  explored at the LHC to find a hint for physics beyond the standard model, though no clear signals have been found yet.

While signals for new physics are actively being probed at LHC, the lower limit of new particle masses  has been pushed up to almost $2~{\rm TeV}$ at the Run 2 of LHC~\cite{Aaboud:2017yyg,CMS:2016crm}, putting most models of physics beyond the standard model (BSM) such as walking technicolor, composite Higgs or supersymmetry in great tension with LHC. We might therefore need to seek alternative solutions to the naturalness problem of the standard model, one of the basic guiding principles for new physics.  
Recently there has been proposed an interesting mechanism to select the Higgs mass dynamically without introducing new physics at the electroweak scale~\cite{Graham:2015cka}. The idea is to construct a model that has many (or infinite) local minima for a  wide range of a field that cosmologically relaxes into a local minimum at the electroweak scale,  starting from a local minimum at the ultraviolet (UV) scale of the Higgs sector, to give a small mass of the electroweak scale to Higgs fields. The QCD axion fits this criterion, if it couples to the Higgs sector, since its potential is periodic under the shift symmetry to have infinitely many local minima, and hence the field is called relaxion. 

In this paper we propose a very minimal model which assumes only very light dilaton in addition to the standard model particles up to a UV cutoff scale, $M$, much higher than the electroweak scale. Our model provides the naturally light Higgs boson, though its UV scale is much higher than the electroweak scale. To discuss the mechanism for our model, we first assume that  our model is an effective theory below the cutoff scale, $M$. One possible candidate for the UV completion of our model, as discussed later, is a dilaton-assisted composite Higgs model, based on Banks-Zaks gauge theories with a quasi infrared (IR) fixed point~\cite{Banks:1981nn}, where both the Higgs boson and the dilaton are (composite) Nambu-Goldstone bosons from strong dynamics in UV, corresponding to the spontaneously-broken global symmetry~\cite{Kaplan:1983fs} and scale symmetry, respectively.     
Being a Nambu-Goldstone boson, associated with spontaneously-broken scale symmetry at the UV scale of the Higgs sector, the dilaton in our model does a similar role as relaxion that alleviates the naturalness problem of the standard model Higgs.  

The standard model is scale invariant classically, if one turns off the Higgs mass or the relevant operators in the Higgs potential. In a classic paper~\cite{Coleman:1973jx}, however, Coleman and Weinberg (CW) showed that, even if one imposes the scale invariance at the quantum level in the Higgs sector of SM,  the Higgs field could develop a vev to break the electroweak symmetry spontaneously by the radiative corrections. Since the value of Higgs vev is determined by the dimensional transmutation of the quartic coupling in the CW mechanism, it should be chosen by experiments; $\left<\phi\right>=v_{\rm ew}\simeq 246~{\rm GeV}$ to account for the weak interactions. The standard model fermions and the weak gauge bosons get mass from the Higgs vev through the Yukawa and gauge couplings with the Higgs fields. The problem of CW mechanism is however that the Higgs mass turns out to be too small, compared to the experimental value, $m_{H}\simeq 125~{\rm GeV}$, unless one introduces extra bosons~\cite{Dermisek:2013pta,Hill:2014mqa}. Furthermore, the standard model has to be fine-tuned from the intrinsic ultraviolet scales such as the Landau pole associated with the weak hypercharge to keep the scale invariance~\cite{Bardeen:1995kv}. 
Our model relies on the electroweak symmetry breaking  {\it {\` a} la} Coleman-Weinberg but evades these problems by embedding the Higgs sector into an almost stable conformal sector at the UV scale of the standard model, which leads to a very light dilaton that generates additional contributions to the Higgs mass of the order of the Higgs vev, 
$\left<\phi\right>=v_{\rm ew}$.

The ultraviolet theory of the Higgs sector in the standard model  is assumed to be near conformal such as the gauge theories with the Banks-Zaks infrared (quasi) fixed point~\cite{Banks:1981nn} and the scale symmetry is spontaneously broken near the IR fixed point to generate a very light dilaton as a Nambu-Goldstone boson. The dilaton of the UV sector then drives the Higgs mass to a small value, controlled by the scale anomaly or the vacuum energy of the UV sector, once the Higgs field develops a vev.   
At low energy our model contains only the standard model with very light dilaton, which is therefore different from previous models~\cite{Bardeen:1995kv,Meissner:2006zh} that attempt to solve the naturalness problem by imposing the scale invariance in the Higgs sector, not broken spontaneously. We also show that the light dilaton of our model abundantly constitutes the dark matter in our universe once it is non-thermally produced at early universe by the vacuum misalignment of the dilaton field. Finally we propose a specific dilaton-assisted composite Higgs model to realize our scenario that the very light dilaton derives the Higgs mass to the electroweak scale.

\section{Very light dilaton and scale anomaly}
The standard model (SM) of elementary particles is scale-invariant in the classical limit, if one turns off the Higgs mass term (and also the cosmological constant term, which we neglect in our discussions), but the scale symmetry is broken radiatively by quantum effects.
Since our model assumes a spontaneously-broken scale symmetry in the UV theory of the Higgs sector,  one is led at low energy to an extension of the standard model that still preserves the scale symmetry at the operator level up to the scale anomaly, though spontaneously broken.

\subsection{Coleman-Weinberg potential.}
We first review the (unsuccessful) scenario of Coleman and Weinberg~\cite{Coleman:1973jx} that Higgs field might be a dilaton in the standard model. 
CW showed that even if one imposes in the standard model the scale-invariance by taking the quadratic term in the Higgs effective potential to vanish
\begin{equation}
\left.\frac{\partial^2V(\phi)}{\partial\phi^2}\right|_{\phi=0}=0\,,
\end{equation}
the scale symmetry is spontaneously broken by radiative corrections. At one-loop, for instance, the Higgs field develops an effective potential to have a minimum away from the origin~\cite{Coleman:1973jx,Gildener:1976ih},
\begin{equation}
V(\phi)=\frac{3}{1024\pi^2}\left[2g^4+(g^2+{g^{\prime}}^2)^2\right]\phi^4\left[\ln\left(\frac{\phi^2}{v_{\rm ew}^2}\right)-\frac12\right],
\label{cw}
\end{equation}
where $g$ and $g^{\prime}$ are the couplings of $SU(2)_L\times U(1)_Y$ electroweak gauge interactions, respectively~\!\footnote{CW did not include top quark. If one includes the top quark, the effective potential changes its sign and the electroweak symmetry does not break. But, one could break it radiatively by introducing new heavy bosons that couple to Higgs fields~\cite{Hill:2014mqa}.}. By expanding the potential around the minimum, one finds the Higgs mass to be 
$m_H^2=(6M_W^4+3M_Z^4)/(8\pi^2v_{\rm ew}^2)\approx (9.8~\!{\rm GeV})^2$ for $v_{\rm ew}=246~{\rm GeV}$ and the weak gauge boson masses, $M_W=80.4~{\rm GeV}$ and $M_Z=91.2~{\rm GeV}$. The scale-invariant Coleman-Weinberg potential leads to too small Higgs mass, compared to the measured value, $125~{\rm GeV}$. Furthermore, if one includes the top quark, the one-loop effective potential changes its sign and the CW mechanism does not work.  As we show however in our model, where the scale symmetry is spontaneously broken at the ultraviolet scale of the Higgs sector, a quadratic term in the Higgs potential is induced at the electroweak scale to generate  the Higgs mass of the electroweak scale, when the dilaton develops a small vacuum expectation value, similar to the relaxion mechanism, provided that the CW mechanism works, having extra bosons~\cite{Hill:2014mqa}.

\subsection{A model for light dilaton }
\label{dilaton-higgs}
Below the UV scale of the standard model, which we denote $M$, taken to be much larger than $1~{\rm TeV}$, the Higgs potential is given as, neglecting possible irrelevant operators,
\begin{equation}
V_0(\phi)=M^2\phi^{\dagger}\phi+\lambda\left(\phi^{\dagger}\phi\right)^2\,
\label{higgs_effective1}
\end{equation}
where the quadratic term is not protected in general and naturally of order of the UV scale, $M$.\footnote{In the composite Higgs model the quadratic term will be quite smaller than $M$, since it is protected by the shift symmetry, broken only radiatively.} Being parameters of the low-energy effective theory, the Higgs mass $M$ and the quartic coupling $\lambda$ include all the ultraviolet contributions from the UV theory above the cutoff scale that are relevant at low energy. Especially the mass term includes the contributions from the massive modes in the UV theory or certain intrinsic scales of the UV sector such as the scale for the conformal phase transition in the case of conformal UV theories~\cite{Tavares:2013dga}. 
In the case of the composite Higgs model, that we will focus on later as a possible model that realizes our mechanism, the Higgs mass is protected by the shift symmetry and generated by the Higgs interactions with the standard model particles such as top quark or EW gauge bosons, the mass term in eq.~(\ref{higgs_effective1}) then should be regarded as the counter term to the SM contributions to the Higgs mass that contains the effect of UV physics.

\begin{figure}
        \begin{minipage}[b]{3in}
\centering%
            \includegraphics[width=2.6in]{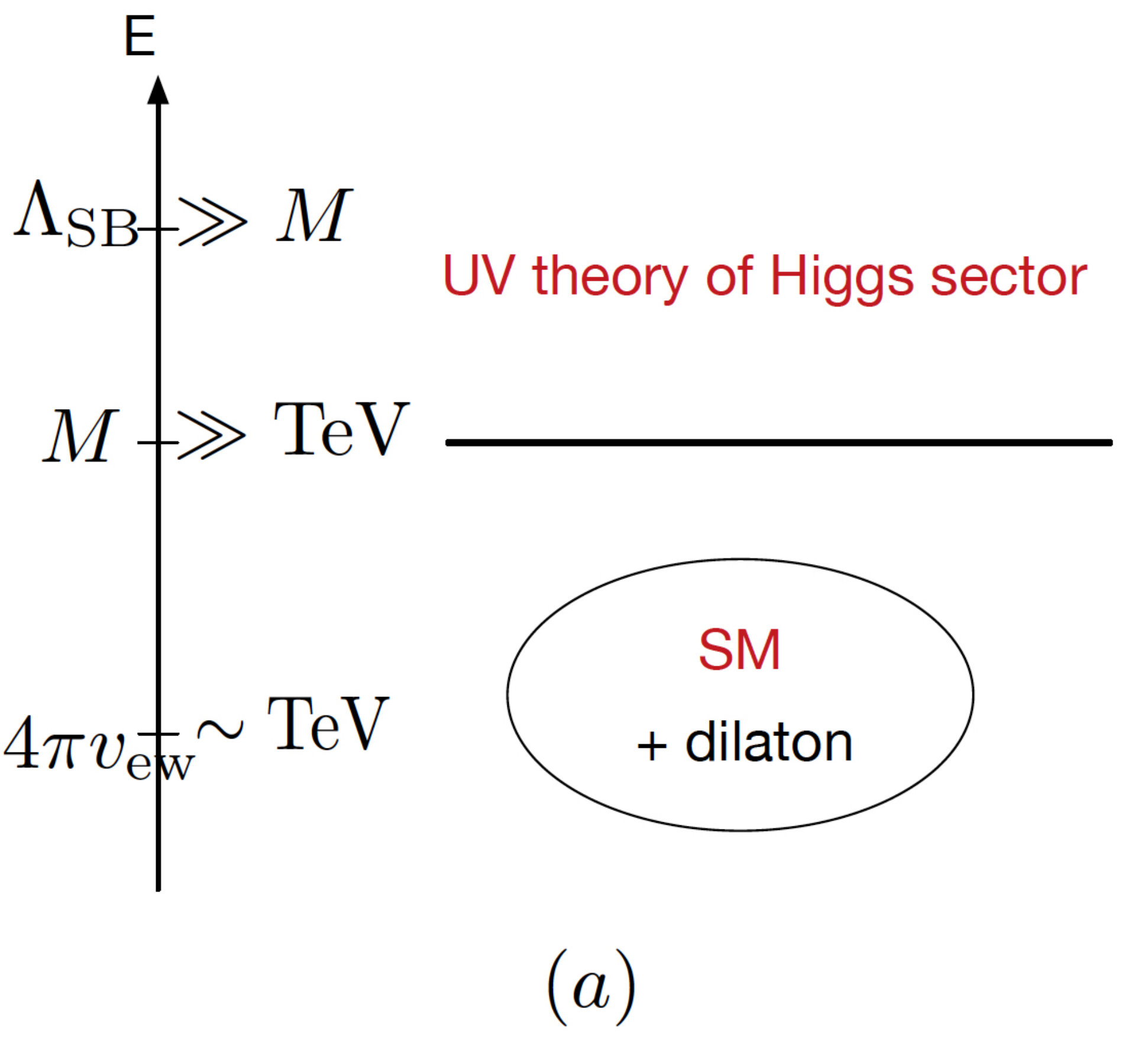}
        \end{minipage}
        \hskip 0.2in
        \begin{minipage}[b]{3in}
\centering%
        \includegraphics[width=3in]{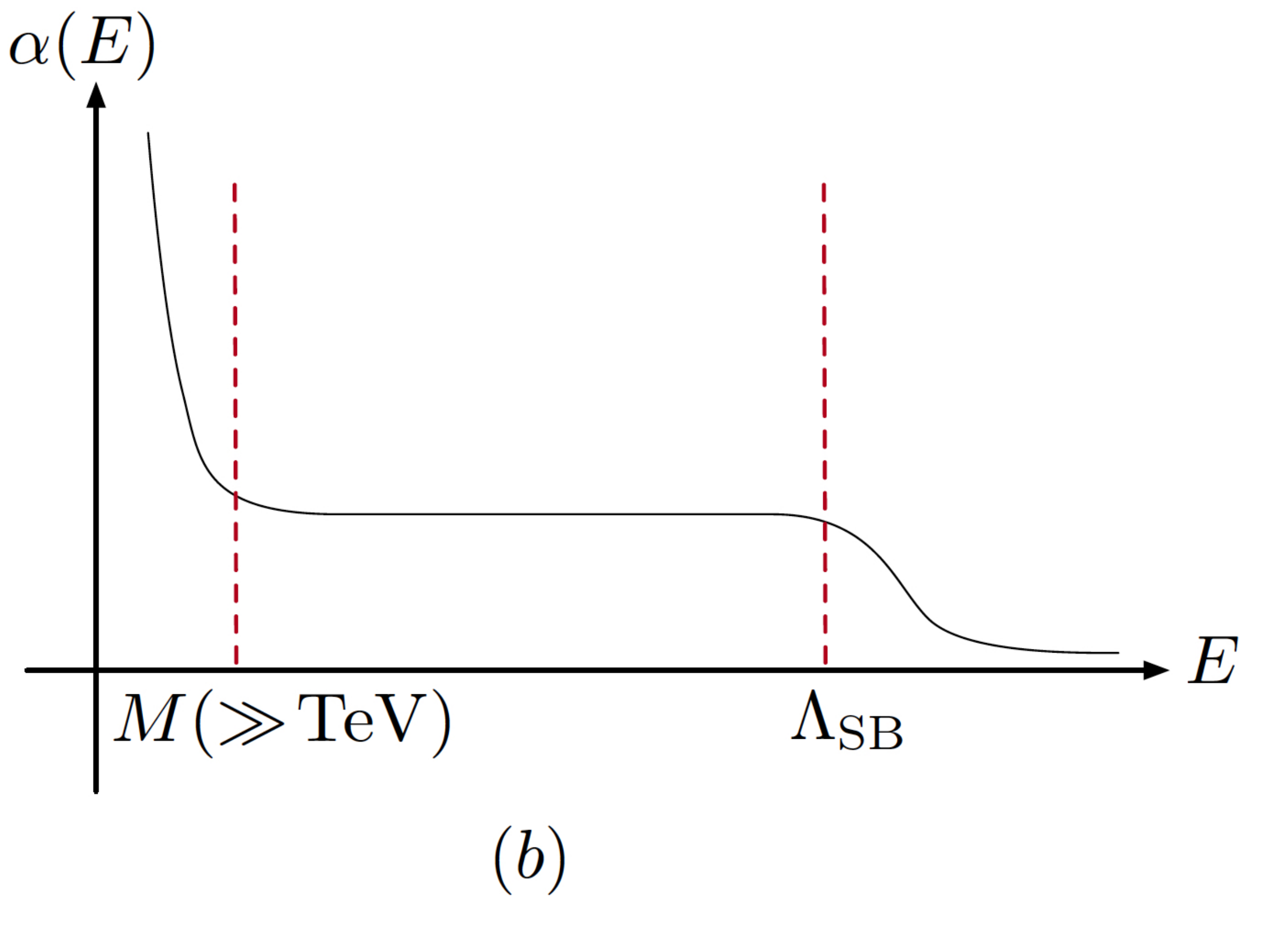} 
       \end{minipage}
\caption{(a) A scale-invariant UV theory of the SM Higgs sector above $M\gg 4\pi v_{\rm ew}$, where the scale symmetry is spontaneously broken at $\Lambda_{\rm SB}\gg M$. Below $M$, there is only one extra particle, the dilaton, in addition to the SM particles. (b) The behavior of couplings in the UV theory as a function of scale, $E$.}   
\label{uv_higgs}
\end{figure}

Since the scale symmetry is assumed to be spontaneously broken near the infrared fixed point of the UV theory like the Banks-Zaks theories, the symmetry breaking scale is much higher than the dynamical or infrared scale of the UV theory, $\Lambda_{\rm SB}\gg M$, known as Miransky scaling~\cite{Miransky:1996pd} or Berezinskii-Kosterlitz-Thouless (BKT) scaling~\cite{Berezinsky:1970fr,Kosterlitz:1973xp}.   Our UV model is therefore almost scale-invariant for the wide range of scales, $M<E<\Lambda_{\rm SB}$. 
(See figure~\ref{uv_higgs}.)

When the scale symmetry is spontaneously broken, the dilatation current creates a Nambu-Goldstone boson, the dilaton, denoted as $\sigma$, out of vacuum:
\begin{equation}
\left<0\right|{\cal D}_{\mu}(x)\left|\sigma(p)\right>=ifp_{\mu}e^{-ip\cdot x}\,,
\end{equation}
where $f$ is the dilaton decay constant, $f\sim\Lambda_{\rm SB}$ and the dilatation current ${\cal D}_{\mu}=\theta_{\mu\nu}x^{\nu}$ with the energy-momentum tensor $\theta_{\mu\nu}$ that couples to gravity~\footnote{As supported by the Schwinger-Dyson analysis~\cite{Hashimoto:2010nw}, one can see that the dilaton decay constant should be the UV scale or ${\Lambda}_{\rm SB}$ not the IR scale, $M$,   that measures the strength of the scale anomaly, since the dilaton has to decouple from the theory if one takes $M\to0$.}.
In order for the dilaton to behave like the relaxion, it has to couple to the Higgs fields. One natural way to achieve this is to assume that both the dilaton and the Higgs boson come from the same UV dynamics.  
Being the low-energy effective theory of a scale-invariant UV theory of the Higgs sector, all the scale-symmetry violating terms in the Higgs sector are coupled to the dilaton field.  
The (anomalous) Ward identity of the scale symmetry fixes how the dilaton couples to the Higgs fields:
Consider the following Green's function,
\begin{equation}
\left<0\right|T\left\{{\cal D}_{\mu}(x)\phi^{\dagger}\phi(0)\right\}\left|0\right>\,.
\end{equation}
Upon integrating over all spacetime points, after taking the total divergence, one gets
\begin{eqnarray}
0&=&\int{\rm d}^4x\,\partial^{\mu}\left<0\right|T\left\{{\cal D}_{\mu}(x)\phi^{\dagger}\phi(0)\right\}\left|0\right>\\
&=&\int{\rm d}^4x\left<0\right|\left[{\cal D}_{0}(x),\phi^{\dagger}\phi(0)\right]\delta(x_0)\left|0\right>
+\int{\rm d}^4x\left<0\right|T\left\{\theta_{\mu}^{\mu}(x)\phi^{\dagger}\phi(0)\right\}\left|0\right>\,.\label{pcdc}
\end{eqnarray}
If one assumes the second term in eq.~(\ref{pcdc}) is saturated at low energy by the dilaton, known as the hypothesis of partially conserved  dilatation currents (PCDC), then one gets  
\begin{equation}
2\left<0\right|\phi^{\dagger}\phi\left|0\right>\approx f\left<\sigma(0)\right|\phi^{\dagger}\phi\left|0\right>\,,
\end{equation}
which shows that the strength to emit the dilaton by $\phi^{\dagger}\phi$ is $2/f$ as realized in the effective theory by $\frac{2}{f}\sigma\phi^{\dagger}\phi$, the first nontrivial term in the expansion of the nonlinear coupling of the dilaton to the quadratic Higgs fields, $e^{2\sigma/f}\phi^{\dagger}\phi$.

The Higgs sector of the standard model now becomes at low energy ($E<M$), suppressing the Higgs couplings to fermions,
  \begin{equation}
  {\cal L}_H=\frac12e^{2\sigma/f}\partial_{\mu}\sigma\partial^{\mu}\sigma+\left(D_{\mu}\phi\right)^{\dagger}
  \left(D^{\mu}\phi\right)-V({\phi,\sigma}),
  \label{higgs_effective}
  \end{equation}
where $\phi$ is the Higgs field and $D_{\mu}$ is the electroweak covariant derivative. 
The potential $V(\sigma,\phi)$ in the effective theory contains the scale anomaly term $V_A$ and  the Higgs potential term $V_0$ with its coupling to the dilaton, 
\begin{equation}
V(\sigma,\phi)=M^2\,e^{2\sigma/f}\,\phi^{\dagger}\phi+\lambda\left(\phi^{\dagger}\phi\right)^2\,+V_A(\sigma)\,.
\label{pot}
\end{equation}
We note that because the dilaton transforms nonlinearly under the scale transformation in the SM sector, $\sigma\to \sigma +\sigma_0$, $\int{\rm d}^4x \,M^2e^{2\sigma/f}\phi^{\dagger}\phi$ is scale invariant and the scale anomaly term changes accordingly, ${\cal E}_{\rm vac}\to {\cal E}_{\rm vac}e^{4\sigma_0/f}$.\footnote{The anomalous Ward identity with $\theta_{\mu}^{\mu}=4{\cal E}_{\rm vac}(\chi/f)^4$ and $\chi=fe^{\sigma/f}$
\begin{equation}
\partial_{\mu}{\cal D}^{\mu}=\theta_{\mu}^{\mu}=4V_A-\chi\frac{\partial V_A}{\partial \chi}\,,
\end{equation}
determines the dilaton potential $V_A(\sigma)$.}

The scale anomaly term in the potential 
is determined by the anomalous Ward identity of scale symmetry as~\cite{Migdal:1982jp}
\begin{equation}
V_A(\sigma)=\left|{\cal E}_{\rm vac}\right|\,e^{4\sigma/f}\left(\frac{4\sigma}{f}-1\right)\,,
\label{anomaly}
\end{equation}
where ${\cal E}_{\rm vac}\sim M^4$ is the vacuum energy density of the UV theory of the Higgs sector~\footnote{The vacuum energy ${\cal E}_{\rm vac}$ in eq.~(\ref{anomaly}), that contributes to the dilaton mass, is due to the vev of the order parameter of the scale symmetry, subtracting out the usual perturbative contributions, so that it vanishes when the vev vanishes~\cite{Miransky:1989qc}.} and $f$ is the dilaton decay constant, $f\gg M$.

The low energy theorem associated with the scale anomaly determines the dilaton mass, $m_D^2=16\left|{\cal E}_{\rm vac}\right|/f^2$. As long as the scale symmetry is broken very close to the (quasi) infrared fixed point of the UV theory, there will be a large separation of two scales $f\sim\Lambda_{\rm SB}$ and $M$, the dynamical (or infrared) scale of the (quasi) scale-invariant UV theory. We then have  $\left|{\cal E}_{\rm vac}\right|\sim M^4\ll f^4$ and the dilaton can be very light~\cite{Choi:2011fy,Choi:2012kx}.

\begin{figure}
\centering%
            \includegraphics[width=4in]{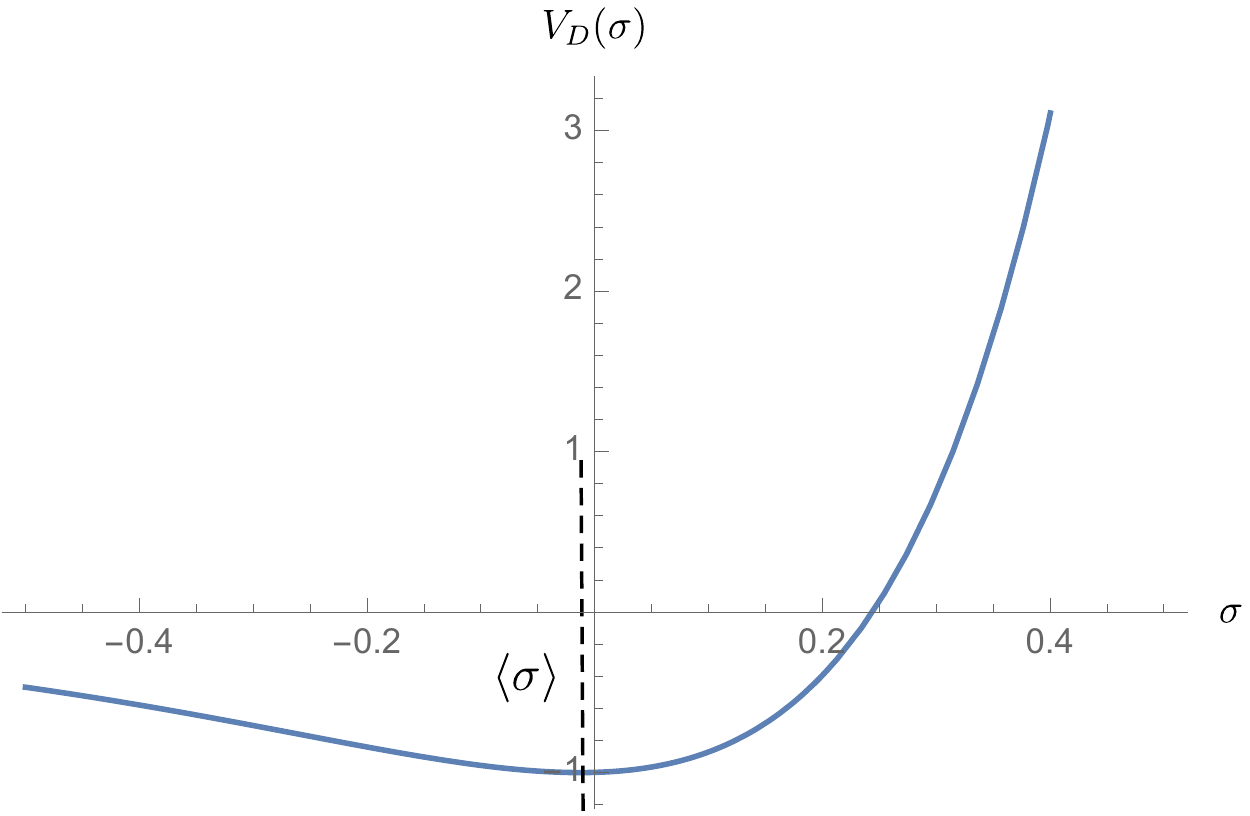}
\caption{The dilaton potential in arbitrary units. The dilaton gets a vev away from the origin once the Higgs develops the electroweak vev.}   
\label{dilaton_pot}
\end{figure}

Since the UV completion of the Higgs sector is assumed to be (quasi) scale-invariant, one can impose the scale invariance at the cutoff scale on the standard model in the sense of Bardeen's naturalness~\cite{Bardeen:1995kv}\footnote{Our renormalization condition at the cutoff scale is technically different from that of Bardeen's proposal.}. We therefore choose the renormalization condition or the counter terms in eq.\,(\ref{pot}) such that the quadratic term of the Higgs field vanishes in the full 1PI effective potential~\cite{Coleman:1973jx}: 
\begin{equation}
m^2_{\phi}\equiv\left.\frac{\partial^2V_{\rm eff}}{\partial\phi^{\dagger}\partial\phi}\right|_{\phi=0=\sigma}=0\,.
\label{quadratic term}
\end{equation}
This renormalization process is stable under any UV contributions because the very light dilaton, that coupled to Higgs fields, enjoys the shift symmetry, $\sigma\to \sigma+\sigma_0$. (See more on this in appendix \ref{a}.) The effective potential then becomes
\begin{equation}
V_{\rm eff}(\sigma,\phi)=M^2\left(e^{2\sigma/f}-1\right)\phi^{\dagger}\phi+V_{\rm CW}(\phi)+V_A(\sigma)\,,
\label{effective potential}
\end{equation}
where $V_{\rm CW}(\phi)$ is the Coleman-Weinberg potential for (massless) Higgs fields. 
At one-loop
\begin{equation}
V^{\rm 1-loop}_{\rm CW}(\phi)=\lambda\left(\phi^{\dagger}\phi\right)^2+\frac18\beta\left(\phi^{\dagger}\phi\right)^2\left[\ln\left(\frac{\phi^{\dagger}\phi}{M^2}\right)-a\right]\,,
\label{one-loop}
\end{equation}
where $a$ is a constant, to be chosen such that $\left<\phi\right>=v_{\rm ew}$, and $\beta$ is nothing but the one-loop beta function of the Higgs quartic coupling, $\lambda$, assumed to be positive by having extra bosons~\cite{Hill:2014mqa}. 
As the Higgs sector flows into the infrared,  the Higgs field develops a vev by the CW mechanism~\cite{Coleman:1973jx}. As soon as  the Higgs field gets a vev, it drives the minimum of the dilaton potential away from the origin, $\left<\sigma\right>\ne0$. 
When the Higgs field develops a vev, $\left<\phi\right>=v_{\rm ew}$, it breaks the scale symmetry explicitly and the dilaton potential gets an additional contribution (See figure~\ref{dilaton_pot})
\begin{equation}
V_D(\sigma)=V_A(\sigma)+V_{\rm CW}(v_{\rm ew})
+M^2\left(e^{2\sigma/f}-1\right)v_{\rm ew}^2\,,
\end{equation}
where $V_{\rm CW}(v_{\rm ew})$ now depends on $\sigma$ from the minimization of $V(\sigma,\phi)$. 
The dilaton field therefore develops a vev away from  the origin. For the one-loop CW potential one finds 
\begin{equation}
-\frac{\left<\sigma\right>}{f}\approx \frac{M^2v_{\rm ew}^2}{8\left|{\cal E}_{\rm vac}\right|}\ll1\,,
\end{equation}
where we have taken the vacuum energy, $\left|{\cal E}_{\rm vac}\right|\gg  M^2v_{\rm ew}^2$.
The Higgs mass then becomes, neglecting small mixing with the dilaton,  
\begin{equation}
m_H^2\equiv V^{\prime\prime}\left(\left<\sigma\right>,\phi\right)\left.\right|_{\phi=v_{\rm ew}}\sim\frac{M^4}{\left|{\cal E}_{\rm vac}\right|}v_{\rm ew}^2\,.
\label{higgs_mass}
\end{equation}
Since the dynamical scale or the infrared scale of the UV theory of the Higgs sector is assumed to be of order of $M$, its vacuum energy  $\left|{\cal E}_{\rm vac}\right|\approx c M^4$, where the constant $c$ is given by the structure of the UV theory. In the case of Banks-Zaks gauge theories with a quasi IR fixed point,  the constant  depends only on the gauge group and the number of fermions~\cite{Miransky:1989qc}.  Thus the Higgs mass is naturally given as the electroweak scale or $v_{\rm ew}$. 

In our model, therefore, having the scale-invariant UV theory of the Higgs sector, that gives the coupling between the dilaton and the Higgs boson,  the dilaton dynamically relaxes the Higgs mass to the electroweak scale, giving the naturally light Higgs boson or $m_H\ll M$. Without severe fine-tuning we have therefore dynamically raised the ultraviolet scale of the Higgs sector $M$ to be much higher than the electroweak scale, alleviating the naturalness problem associated with the Higgs mass. 
The scale symmetry does a crucial role in our mechanism. Having the very light dilaton at the UV scale $M$, the Higgs sector is almost scale-invariant.  The curvature of the Higgs potential, therefore, has to be chosen to vanish at the origin by the renormalization condition to be consistent with the scale symmetry, $\sigma\to\sigma+\sigma_0$. However, once the Higgs sector flows into IR, the Higgs field develops a vev, $\left<\phi\right>=v_{\rm ew}$ by the CW mechanism, generating the IR scale. The Higgs vev therefore sets the scale for the Higgs mass.

\section{Very light dilaton as dark matter}
Besides the naturalness problem that we discussed, another strong motivation for physics beyond the standard model is to account for the dark matter that constitutes about 23\% of the total energy of our present universe. According to the current standard big-bang cosmology, cold dark matter with a cosmological constant, so-called the $\Lambda$CDM fits the current observations such as the cosmic microwave background (CMB) best~\cite{Bennett:2012zja,Ade:2013sjv}. A very light dilaton has been shown to be one of the best candidates for the cold dark matter~\cite{Choi:2011fy,Choi:2012kx}.

\subsection{Life time}
\begin{figure}[h!]
\centering   
  \includegraphics[width=5in]{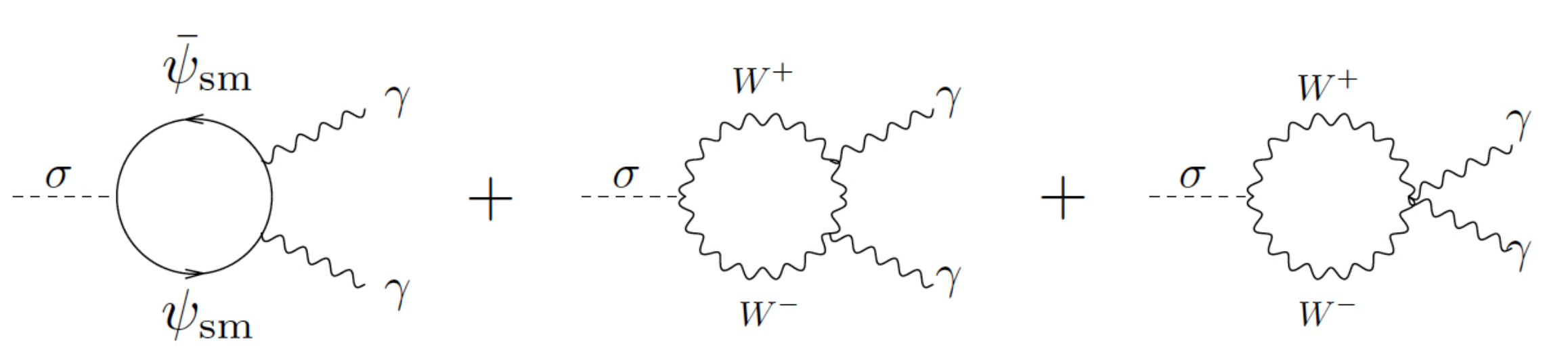}
  \caption{Leading order diagrams that dilaton decays into two photons. The dashed line denotes dilaton and the solid line denotes SM fermions. The external curly lines denote photons and the internal curly lines denote $W$ bosons. }
  \label{fig2}  
\end{figure}

The dilaton couples to the standard model particles, once they get mass by the Higgs mechanism that breaks the electroweak symmetry. The light dilaton therefore decays into two photons through a loop process (and also into neutrinos and gravitons, which we neglect), as shown in figure~\ref{fig2}. 
The decay rate is given at one loop for the very light dilaton as 
\begin{equation}
\Gamma(\sigma\to\gamma\gamma)\simeq \frac{\alpha_{\rm em}^2}{36\pi^3}\frac{m_D^3}{f^2}\left|{\cal C}\right|^2\,,
\end{equation}
where ${\cal C}$ is approximately a constant times the electric charge squared,  summed over all charged particles in the standard model. 
We estimate the lifetime of the dilaton
\begin{equation}
 \tau_{\rm D} \simeq 10^{20} \, {\rm sec}  \, \left(\frac{5}{\cal C}  \right)^2  
 \left( \frac{10 \, {\rm keV}}{m_{\rm D}} \right)^3 \left( \frac{f}{10^{12} \, {\rm GeV}} \right)^2\,.
\end{equation}
In order for the dilaton to be long-lived to become a dark matter candidate of mass, $m_D=10~{\rm keV}$ with decay constant $f=10^{12}~{\rm GeV}$, the UV scale has to be $M\sim 10~{\rm TeV}$ by the low energy theorem for the dilaton, $m_D^2f^2=16\left|{\cal E}_{\rm vac}\right|\sim M^4$. If the dilaton decay constant is as high as the GUT scale, $f\sim 10^{16}~{\rm GeV}$, the dilaton mass can be as low as $1~{\rm eV}$. 
For $M\sim 100~{\rm TeV}$, we have $f=10^{15}~{\rm GeV}$, if $m_D=10~{\rm keV}$, or $f=10^{16}~{\rm GeV}$, if $m_D=1~{\rm keV}$. Therefore, if the UV scale of SM is around $10-100~{\rm TeV}$, the dilaton mass is about $1~{\rm eV}-10~{\rm keV}$.

\subsection{Relic abundance of dilaton}
Since the dilaton is weakly coupled, it will not be in thermal equilibrium with other particles in early universe, when it is produced.
However, by the vacuum misalignment the light dilaton will be non-thermally produced in early universe. If we take the degree of the misalignment to be $\theta_{\rm os}=\delta\sigma/f$, the relic density of the dilaton will be at the time of oscillation from the misalignment
\begin{equation}
\rho_{\sigma}(T_{\rm os})=\left|V_D(T_{\rm os})-V_D^{\rm min}\right|\simeq M^4\,{\theta_{\rm os}}^2\,.
\end{equation}
Since the relic density at present is given as 
$\rho_{\sigma}(T_0)=\rho_{\sigma}(T_{\rm os})\cdot \frac{s(T_0)}{s(T_{\rm os})}$\,,
where $s(T)$ is the entropy density at temperature $T$, 
we find the dilaton dark matter contributes to energy of our present universe as~\cite{Choi:2011fy,Choi:2012kx}  
\begin{equation}
\Omega_{\sigma}h^2\sim 0.5\!\left(\frac{\delta\sigma}{10^{-5}f}\right)^2\!\!
\left(\frac{110}{g_*(T_{\rm os})}\right)\!\!\left(\frac{M}{10{\rm TeV}}\right)^4\!\!\left(\frac{10{\rm TeV}}{T_{\rm os}}\right)^3\,,
\end{equation}
where $g_*(T_{\rm os})$ is the effective degrees of freedom of early universe at the temperature for the coherent dilaton field starting to oscillate. Very light dilaton as dark matter has been studied in~\cite{Choi:2011fy,Choi:2012kx} in the context of walking technicolor. The light dilaton in our model might be detected in similar experiments such as a microwave cavity experiment under strong magnetic fields.

\section{Dilaton-assisted composite Higgs model}
In this section we propose a specific model to realize our scenario that the dilaton relaxes the Higgs mass to the electroweak scale. This model is based on a composite Higgs model, where the Higgs boson is a pseudo Nambu-Golstone boson, associated a global symmetry, broken spontaneously by strong dynamics at $M\gtrsim 4\pi v_{\rm ew}$~\cite{Kaplan:1983fs,Contino:2010rs}. The Higgs mass is protected by the (approximate) shift symmetry that is radiatively broken by the electroweak interactions, giving the loop-suppressed Higgs mass,
\begin{equation}
m_H^2\sim \frac{{\tilde g}^2}{16\pi^2} M^2\,,
\end{equation}
where ${\tilde g}$ is the coupling of  the electroweak interactions.
On top of these features of the composite Higgs, our model needs to exhibit a (quasi) IR fixed point to have a very light dilaton at low energy that couples to the Higgs fields. 

Consider a composite Higgs model based on the ${\rm SU}(2)$ gauge theory with $N_f$ Dirac fermions $\psi^i$ ($i=1,2,\cdots, N_f$) of the fundamental representation~\cite{Galloway:2010bp,Cacciapaglia:2014uja} and with $N_s$ Dirac fermions $\chi^i$ ($i=1,2,\cdots, N_s$) in the symmetric second-rank ternsor representation~\cite{Hong:2004td}. Since the spinors are pseudo real in the ${\rm SU}(2)$ gauge theory, the global symmetry is ${\rm SU}(2n)$ for $n$ (massless) Dirac fermions, which breaks down to ${\rm Sp}(2n)$, once the fermion bilinears form condensates~\cite{Lewis:2011zb}. The Higgs field is then identified as one of the Goldstone bosons living on the coset space, ${\rm SU}(2n)/{\rm Sp}(2n)$, where the SM gauge group is embedded in its unbroken subgroup, $ {\rm SU}(2)_L\times {\rm U}(1)_Y\subset {\rm Sp}(2n)$ so that the Higgs fields transform correctly under the SM gauge symmetry. 

To see whether our composite Higgs model is near the conformal window or not, we study the two-loop beta function of the ${\rm SU}(N)$ gauge theory with $N_f$ fundamental Dirac fermions and $N_s$ Dirac fermions in the second-rank symmetric tensor representation, that is given as
\begin{equation}
\beta(\alpha)\equiv\mu\frac{\partial\alpha}{\partial\mu}=-b\alpha^2-c\alpha^3\,,
\end{equation}
with the coefficient $b$ and $c$, known as 
\begin{eqnarray}
6\pi b&=&11N-2 N_f-2N_s(N+2)\\
24\pi^2c&=& 34 N^2-10NN_f-3\left(N-\frac{1}{N}\right)N_f\nonumber\\&&-10NN_s(N+2)-\frac{6}{N}(N-1)(N+2)N_s(N+2)\,.
\end{eqnarray}
The theory will be asymptotically free if $b>0$ and will have a IR fixed point near at $\alpha_{\ast}=-b/c$, if $c<0$ and the chiral symmetry is unbroken. The chiral symmetry of the Dirac fermions will break at the critical couplings, $\alpha_c(f)$ and $\alpha_c(s)$ for the fermions in the fundamental representation and in the symmetric second-rank tensor, respectively, if they are smaller than the would-be IR fixed point $\alpha_{\ast}$. The critical couplings are given in the ladder approximation~\cite{Cohen:1988sq,Hong:1989zza} as 
\begin{equation}
\alpha_c(f)=\frac{2\pi}{3}\frac{N}{N^2-1},\quad \alpha_c(s)=\frac{2\pi}{3}\frac{N}{(N+2)(N-1)}\,.
\end{equation}

For the ${\rm SU}(2)$ gauge theory with $N_f=8$ fundamental Dirac fermions the lattice results show that the theory is in the conformal window, flowing into a stable IR fixed point~\cite{Leino:2015bfg}. This is consistent with our two-loop beta function analysis, which shows that the critical coupling for the chiral symmetry breaking $\alpha_c(f)=1.40$ is bigger than the IR fixed point, $\alpha_{\ast}\approx1.26$. 
Let us consider another gauge theory in the conformal window; the ${\rm SU}(2)$ gauge theory with $N_f=4$ Dirac fermions in the fundamental representation and $N_s=1$ Dirac fermion in the symmetric second-rank tensor representation. Since the critical couplings for both representations, $\alpha_c(f)=1.40$ and $\alpha_c(s)=1.05$ are larger than the IR fixed point, $\alpha_{\ast}\simeq0.84$, the theory will be in the conformal window, according to the analysis based on the two-loop beta function. The theory will flow from the asymptotically free theory to the IR fixed point. The coupling never becomes strong enough to break the chiral symmetry. Now, we gauge half of the flavor of the fundamental Dirac fermions so that they become bi-fundamental under ${\rm SU}(2)_1\times{\rm SU}(2)_2$ (See Table~\ref{matter}.).
\begin{table}[ht]
\caption{The matter content of the gauge theory near the conformal window.}
\centering
\begin{tabular}{c| c c }
\hline
 & ${\rm SU}(2)_1$ & ${\rm SU}(2)_2$  \\ [0.5ex] 
\hline
$\psi^1_{a\alpha}$ &\ $\fund$ &\ $\fund$ \\
$\psi^2_{a\alpha}$ &\ $\fund$ &\ $\fund$ \\
$\chi_{\{ab\}}$ &\ $\symm$ &\ $1$   \\ [1ex]
\hline
\end{tabular}
\label{matter}
\end{table}

\begin{figure}[ht!]
\centering   
  \includegraphics[width=3in]{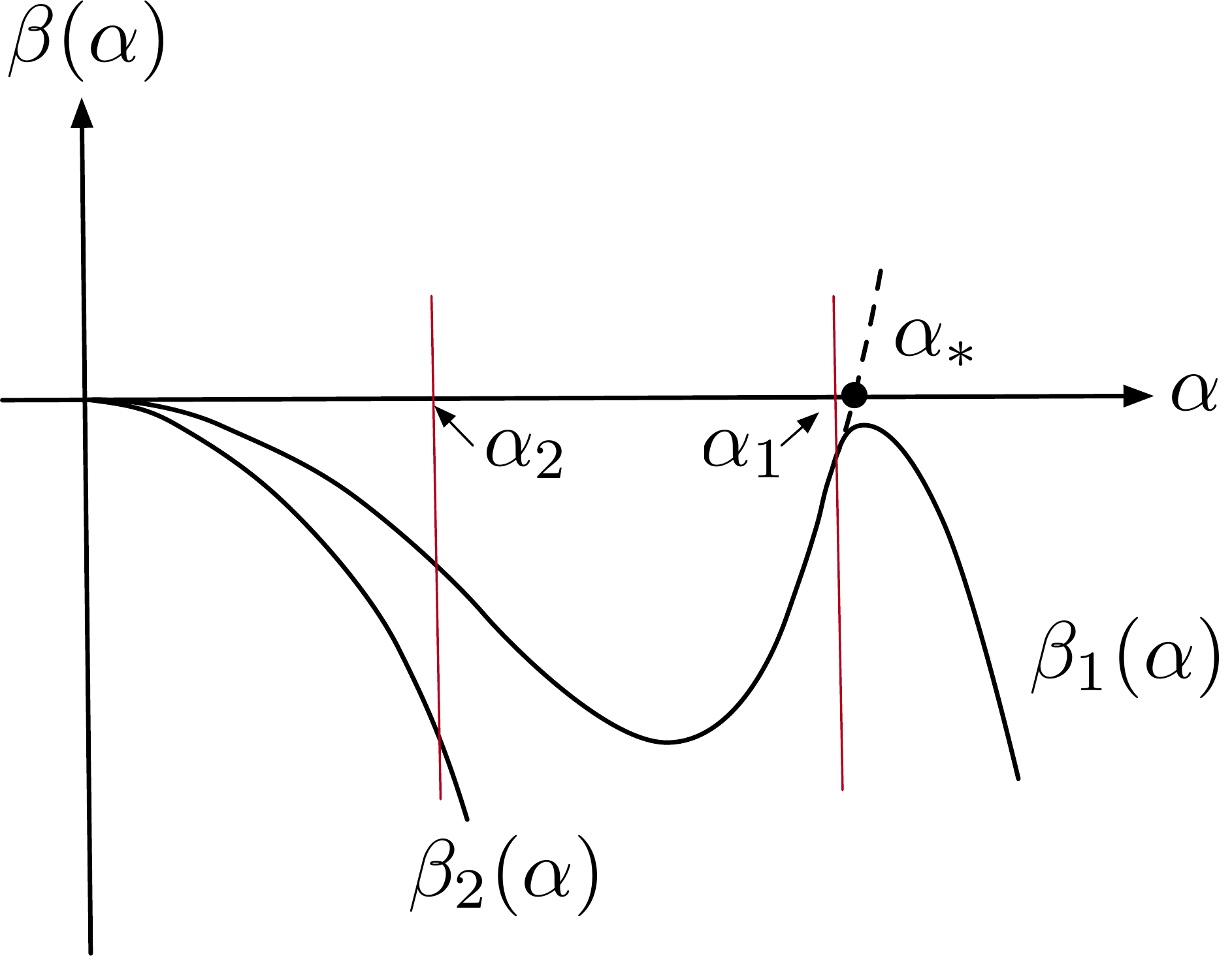}
  \caption{The beta functions $\beta_1$ and $\beta_2$ of the ${\rm SU}(2)_1\times{\rm SU}(2)_2$ gauge theory.  The chiral symmetry of the bi-fundamental Dirac fermions is broken at $\alpha_1\approx\alpha_{\ast}$ for ${\rm SU}(2)_1$ and $\alpha_2\approx\alpha_c(f)-\alpha_{\ast}$ for ${\rm SU}(2)_2$. }
  \label{beta}  
\end{figure}

For the bi-fundamental  fermions $\psi^{i}$ ($i=1,2$) the attractive forces are additive and thus
the critical couplings for the chiral symmetry breaking will be smaller than $\alpha_c(f)=1.40$ in the ladder approximation, since the Bethe-Salpeter Kernel for the fermion-bilinear in the scalar channel is approximately in the short-distance limit~\cite{Hong:1989zza}
\begin{equation}
\frac{\alpha_1+\alpha_2}{\alpha_c(f)}\frac{1}{x^2}\,,
\end{equation}
where $\alpha_i$ is the coupling of ${\rm SU}(2)_i$ at the symmetry breaking scale, $\Lambda_{\rm SB}$, and $x^2$ is the distance square of the four-dimensional Euclidean space. However, unlike the ${\rm SU}(2)_1$ gauge theory, the ${\rm SU}(2)_2$ gauge coupling runs, becoming strong at low energy ($E\ll \Lambda_{\rm SB}$). Therefore we tune $\alpha_2$ to become close to the $\alpha_c(f)-\alpha_{\ast}\approx0.56$ at $E=\Lambda_{\rm SB}$ so that the chiral symmetry of the bi-fundamental fermions breaks  dynamically very near the IR fixed point of the ${\rm SU}(2)_1$ gauge theory~\footnote{Since the bi-fundamental fermions are charged under both gauge groups, the $\beta$-function will have mixings between two gauge couplings. At two-loop $\beta(\alpha)=-b\alpha^2-c\alpha^3+{\tilde b}\alpha^2\alpha_2$ for ${\rm SU}(2)_1$, where ${\tilde b}=3/(8\pi^2)$. The mixing will shift in perturbation the value of $b$ to $b-{\tilde b}\alpha_2$. However, since $\alpha_2$ is at most 0.56 before the chiral symmetry breaking, the mixing does not change the IR fixed point much and thus negligible for our discussions.}~\footnote{We note that by gauging partially the flavor symmetry of the gauge theory, as in our case, one can move most of the gauge theories in the conformal window to the broken phase very near the conformal window.}. Once the bi-fundamental fermions get dynamical mass, they will decouple at low energy and the ${\rm SU}(2)_1$ coupling becomes stronger and stronger to break the ${\rm SU}(2)_{\chi}$ chiral symmetry of $\chi_{ab}$ down to ${\rm U}(1)_{\chi}$ and we will have two extra Goldstone bosons, $\Phi_{\chi}$. By identifying the unbroken ${\rm U}(1)_{\chi}$ as 
the ${\rm U}(1)_{\rm em}$, the Goldstone bosons are oppositely charged and get mass $\sim eM_{\chi}$, where $e$ is the electric charge and $M_{\chi}\sim M$ is the scale for the ${\rm SU}(2)_{\chi}$ chiral symmetry breaking. 
As the ${\rm SU}(2)_1$ gauge theory flows into the IR, the bi-fundamental fermions get condensed at $\Lambda_{\rm SB}$, breaking the chiral symmetry near the (quasi) IR fixed point.  The coupling of ${\rm SU}(2)_1$  will show the walking behavior, since its beta function $\beta_1(\alpha)\approx0$ for the wide range of scales, shown in figure~\ref{beta},
\begin{equation}
M<E<\Lambda_{\rm SB}\,,
\end{equation}
where the dynamical (IR) scale is given by the Miransky or BKT scaling, 
\begin{equation}
M\approx \Lambda_{\rm SB}\exp\left(-\frac{\pi}{\sqrt{{\alpha_{\ast}}/{\alpha_1}-1}}\right)\,.
\label{ir scale}
\end{equation}
We see that the dynamical scale $M$ can be arbitrarily small, compared the chiral symmetry breaking scale $\Lambda_{\rm SB}$, if $\alpha_1$ is close to the IR fixed point $\alpha_{\ast}$. 
Our composite Higgs model therefore is almost scale-invariant for energy $M<E<\Lambda_{\rm SB}$ and there should be a dilaton associated with spontaneous breaking of scale symmetry, when the bi-fundamental Dirac fermions get condensed at $\Lambda_{\rm SB}$ to break its global symmetry ${\rm SU}(4)$ down to ${\rm Sp}(4)$. 

Since the vacuum manifold ${\cal M}={\rm SU}(4)/{\rm Sp}(4)\sim {\rm SO}(6)/{\rm SO}(5)$ is five dimensional, there will be five Goldstone bosons. 
If we embed the standard model gauge group into the unbroken subgroup ${\rm Sp}(4)\sim{\rm SO}(5)\supset {\rm SU}(2)\times {\rm U}(1)$\,, the five Goldstone bosons can be decomposed into one ${\rm SU}(2)_L$ doublet,  to become
the SM Higgs boson, and one real $CP$-odd singlet scalar~\cite{Katz:2005au,Gripaios:2009pe,Cacciapaglia:2014uja}. The broken generator associated with the singlet scalar is nothing but the axial fermion number ${\rm }U(1)_A^{\psi}$ of the bi-fundamental fermion $\psi$. Assuming it is non-anomalous\,\footnote{This is always made possible, if one introduces leptonic fields that are charged under ${\rm U}(1)_A^{\psi}$ but not under the UV gauge interactions. }, we weakly gauge it so that the singlet is absorbed into the ${\rm U}(1)_A^{\psi}$ gauge boson. The ${\rm U}(1)_A^{\psi}$ gauge boson gets mass $\sim g_{\psi}M\gg 4\pi v_{\rm ew}$, with $g_{\psi}$ being the ${\rm U(1)}_A^{\psi}$ coupling, and decouples from the SM particles at low energy.

\begin{figure}[h!]
\centering   
  \includegraphics[width=3.5in]{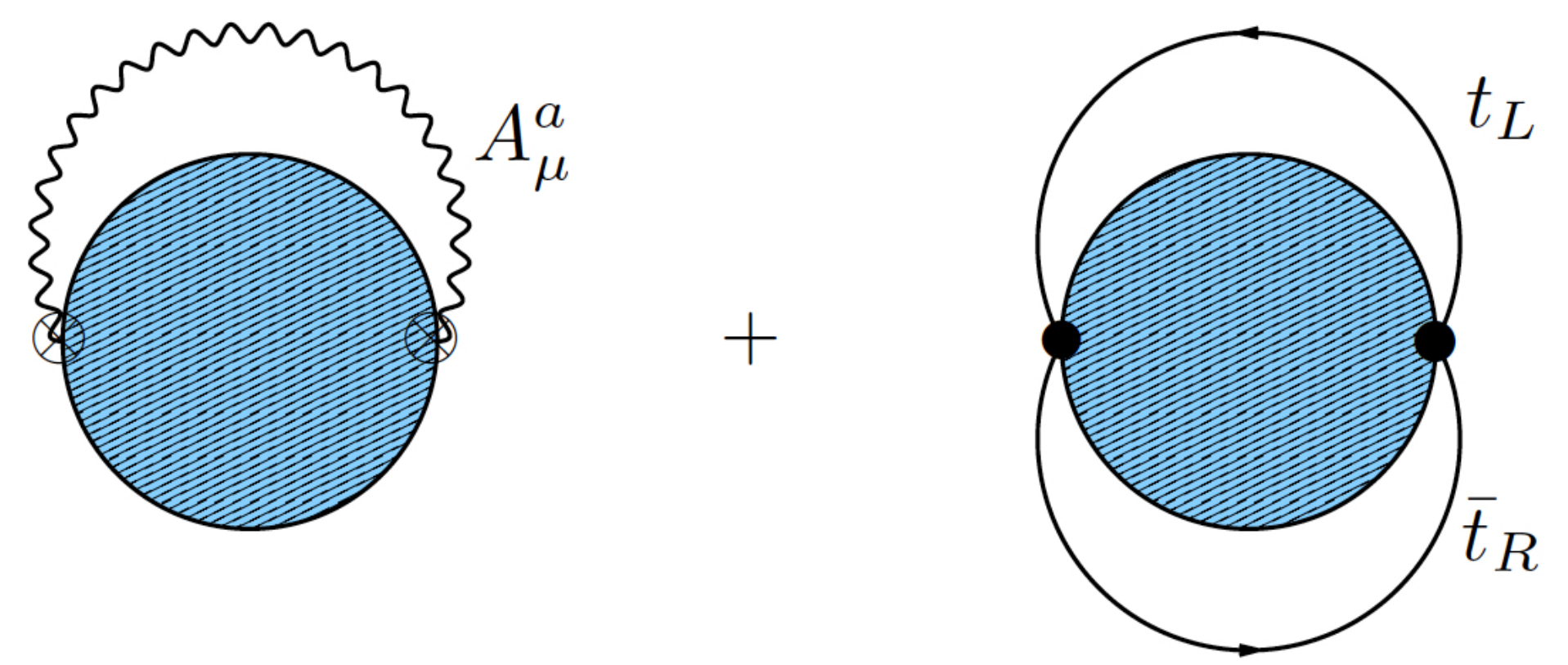}
  \caption{The SM corrections to the vacuum energy. The blob denotes the (full) two-point function of the strong dynamics of the composite Higgs.  The curly line denotes the SM gauge bosons, $A_{\mu}^a$ and the solid line denotes the SM fermions such as the top quark. }
  \label{vac_e}  
\end{figure}

When the ${\rm SU}(2)\times {\rm U}(1)$ subgroup in the unbroken global symmetry is gauged, the electroweak interaction contributes to the vacuum energy, lifting the degeneracy of the vacuum manifold. The correction to the vacuum energy at the leading order in the electroweak coupling expansion is given as (See figure~\ref{vac_e}), after the renormalization, 
\begin{equation}
\Delta {\cal E}_{\rm vac}=-\frac{g_{\rm ew}^2}{2}\int{\rm d}^4x\Delta^{\mu\nu}(x)\left<0\right|U^{\dagger}T\left\{J_{\mu}(x)J_{\nu}(0)\right\}U\left|0\right>\,\equiv\frac{\alpha_{\rm ew}}{4\pi}M^2f_{\phi}^2\,F\left(\frac{\phi}{f_{\phi}}\right)\,,
\label{vac}
\end{equation}
where $\Delta^{\mu\nu}$ is the electroweak gauge boson propagator and $J_{\mu}(x)$ are the electroweak currents, denoted as $\otimes$ in figure~\ref{vac_e}. The composite Higgs field $\phi$ is nonlinearly realized,  $U=\exp\left[2i\phi/f_{\phi}\right]$\, with the decay constant, $f_{\phi}\sim M$ by the Pagels-Stokar formula~\cite{Hashimoto:2010nw}.
In addition to the SM gauge bosons, the SM fermions will contribute to the vacuum energy through the Yukawa interactions. To calculate, for instance, the top Yukawa contributions to the vacuum energy, one needs to calculate the two-point function of the composite operators $\Gamma(x)$ or $\Gamma^{\dagger}(x)$ of the UV theory, denoted as the bullets in figure~\ref{vac_e}, that source or sink the top-quark mass term, connected by the top-quark propagators. The zero mode of the composite operator $\Gamma(x)$ for the top quark should be correctly normalized to give the top Yukawa coupling, $y_t$\,.\,\footnote{The SM fermions are external to the composite Higgs dynamics. Unlike the gauge interactions, the Yukawa interaction of SM fermions will be absent in the composite Higgs, unless  the interaction  for the Yukawa couplings is incorporated in the UV theory to begin with. Here we assume that the Yukawa couplings are generated in the UV theory through the four-Fermi interactions between the SM fermions and the fermions in the UV theory of the composite Higgs, similar to the extended technicolor~\cite{Dimopoulos:1979es,Eichten:1979ah}.}

Expanding the vacuum energy of the composite Higgs due to the vacuum misalignment  in powers of the Higgs fields, $\phi$, one finds the Higgs effective potential at the scale $M$ becomes for $\phi^{\dagger}\phi\ll f_{\phi}^2$
\begin{equation}
V_0(\phi)=M_0^2\phi^{\dagger}\phi+\frac{\beta}{8}\left(\phi^{\dagger}\phi\right)^2\left[\ln\left(\frac{\phi^{\dagger}\phi}{M^2}\right)-a\right]+\cdots\,,
\end{equation}
where $M_0^2=\xi M^2$ with $\xi\approx 3g_{\psi}^2/32\pi^2-3y_t^2/4\pi^2$.  The one-loop beta-function for the Higgs quartic coupling $\beta$ is adjusted to be positive in the composite Higgs model. For instance, the ${\rm U}(1)_A^{\psi}$ gauge-boson contribution to the one-loop beta-function to the quartic coupling is given as 
\begin{equation}
\beta_1=\frac{3}{32\pi^2}g_{\psi}^4\,,
\end{equation}  
which makes the beta-function $\beta>0$ as long as $g_{\psi}\gtrsim2y_t$.

In the dilaton-assisted composite Higgs model  the (one-loop) effective potential for the composite Higgs fields and the dilaton is given as 
\begin{equation}
V(\sigma,\phi)=M_0^2\left(e^{2\sigma/f}-1\right)\phi^{\dagger}\phi+\frac{\beta}{8}\left(\phi^{\dagger}\phi\right)^2\left[\ln\left(\frac{\phi^{\dagger}\phi}{M^2}\right)-a\right]+V_A(\sigma)\,,
\end{equation}
where we have chosen  the renormalization condition that is consistent with the scale symmetry~\cite{Bardeen:1995kv}, 
\begin{equation}
\left.\frac{\partial^2V}{\partial\phi^{\dagger}\partial\phi}\right|_{\sigma=0=\phi}=0\,.
\label{condition}
\end{equation}
To find the vacuum configuration we minimize the effective potential:
\begin{equation}
\left.\frac{\partial V}{\partial\sigma}\right|_{\phi=v_{\rm ew}}=M_0^2\frac{2}{f}e^{2\sigma/f}v_{\rm ew}^2+\left|{\cal E}_{\rm vac}\right|\frac{16\sigma}{f^2}e^{4\sigma/f}=0,
\end{equation}
which gives
\begin{equation}
-\frac{\left<\sigma\right>}{f}\approx\frac{M_0^2v_{\rm ew}^2}{8\left|{\cal E}_{\rm vac}\right|}=\frac{M_0^2v_{\rm ew}^2}{8cM^4}\ll1\,,
\end{equation}
using the relation ${\cal E}_{\rm vac}=-cM^4$ of the composite Higgs model.
Neglecting the small mixing with the dilaton, the Higgs mass becomes
\begin{equation}
m_H^2=\left.\frac{\partial^2}{\partial\phi^{\dagger}\partial\phi}V\left(\left<\sigma\right>,\phi\right)\right|_{\phi=v_{\rm ew}}=\left(\frac{\xi}{4c}+\frac{\beta}{4}\right)v_{\rm ew}^2\,.
\end{equation}
Since in our composite Higgs model $c\simeq 1.2$~\cite{Miransky:1989qc},  either $\xi$ or $\beta$ has to be ${\cal O}(1)$ or the ${\rm U}(1)_A^{\psi}$ coupling $g^2_{\psi}/4\pi \simeq 0.73$ to give $m_H\simeq 125~{\rm GeV}$. 

By coupling the Higgs sector to the light dilaton, we have shown that  the Higgs mass is given by the IR scale, $m_H\sim v_{\rm ew}$, not by the UV scale, $M$. This seems mysterious but the scale symmetry is working behind. As the Higgs sector flows into IR, $M\to M^{\prime}$, the dilaton transforms $\sigma\to \sigma+f\ln\left(M^{\prime}/M\right)$ to keep the renormalization condition (\ref{condition}) until the Higgs field gets the vev, $\left<\phi\right>=v_{\rm ew}$ which breaks the scale symmetry. 
Hence the UV scale of the composite Higgs can be arbitrarily high. The cosmological or phenomenological requirements on the dilaton mass and its decay constant, however, will constrain the scale of the model. In our model with the ${\rm SU}(2)_1\times{\rm SU}(2)_2$ composite-Higgs gauge group, if we take for instance $M=10~{\rm TeV}$ and $\alpha_1=0.98\,\alpha_{\ast}$,  the dilaton decay constant $f\sim\Lambda_{\rm SB}\simeq 3\times 10^{10}~{\rm TeV}$ to give the dilaton mass
\begin{equation}
m_D\sim \frac{M^2}{f}\simeq 3~{\rm keV}\,.
\end{equation}
The dilaton of this mass range  is shown to be a good candidate for the dark matter~\cite{Choi:2011fy,Choi:2012kx}.

\section{Discussions and conclusion}

In this paper we propose a mechanism that very light dilaton naturally derives the Higgs mass to the electroweak scale, if the Higgs field gets the electroweak vev {\it{\`a la}} Coleman-Weinberg mechanism and couples to the light dilaton. 
The scale symmetry, associated with the light dilaton, does a crucial role in our mechanism that the Higgs mass is given by the Higgs vev, $v_{\rm ew}$, the IR scale of the Higgs sector.

We then show that the dilaton-assisted composite Higgs model, based on the ${\rm SU}(2)_1\times {\rm SU}(2)_2$ gauge theory with two Dirac fermions in the bi-fundamental representation and one in the symmetric tensor representation of ${\rm SU}(2)_1$, realizes our scenario. Both the dilaton and the composite Higgs are shown to arise as (pseudo) Nambu-Goldstone bosons, once the Dirac fermions in the bi-fundamental representation get condensed.
The standard model is then coupled through the very light dilaton to the quasi-conformal composite Higgs model at $M\gg1~{\rm TeV}$. By imposing the scale symmetry on the standard model, the naturalness problem of Higgs mass is alleviated to the UV scale, $M$. 
When the electroweak symmetry is radiatively broken by the CW mechanism, the dilaton potential gets an extra contribution from the Higgs vev, which then drives the dilaton vev away from the origin. 
The non-vanishing dilaton vev relaxes the Higgs mass naturally to be of the electroweak scale, as the vacuum energy or the scale anomaly of the scale-invariant UV theory of the Higgs sector is of the UV scale, $M$~\footnote{Our mechanism that the light dilaton relaxes the Higgs mass from the UV scale $M$ to the IR scale, $v_{\rm ew}$ might be the manifestation of the $a$-theorem in the conformal field theory, studied in~\cite{Komargodski:2011vj}. This needs to be investigated further.}.   

At the electroweak scale, much below the UV scale,  the model contains the standard model and only one extra particle, the very light dilaton, which is shown to be a good candidate for dark matter in the universe. If we take for instance the UV scale $M\sim10-100~{\rm TeV}$ and the dilaton decay constant $f\sim 10^{12-16}~{\rm GeV}$, the dilaton mass becomes $m_D\sim 1~{\rm eV}-10~{\rm keV}$, which is then long lived enough and abundantly produced by the vacuum misalignment to constitute dark matter in our universe.  

Finally, the dilaton-assisted composite Higgs model predicts in addition to the very light dilaton a heavy (axial) vector boson of mass $\sim g_{\psi}M$ and two, oppositely charged, pseudo Nambu-Goldstone bosons (SM singlet) of mass $\sim eM$. If the UV scale of our composite Higgs model is around a few $10~{\rm TeV}$, their mass will be a few TeV or so, accessible at LHC.

\acknowledgments
The author thanks D. Bak, K.\,Y. Choi,  G. Giudice, D.\,B. Kaplan, D. Kim, H.\,D. Kim, J. Serra, A. Strumia and A. Wulzer for useful comments and discussions. The author is also grateful to the CERN Theory group for the hospitality during his participation at the CERN-CKC TH institutes and visit to CERN, where part of this work was done. 
This research was supported by Basic Science Research Program through the National Research Foundation of Korea (NRF) funded by the Ministry of Education (NRF-2017R1D1A1B06033701).

\appendix
                                  
\section{On the stability of the renormalization condition $m_{\phi}^2(\Lambda)=0$}
\label{a}
In this appendix we show that the renormalization condition, imposed in eq.\,(\ref{effective potential}), that the Higgs quadratic term vanishes at the UV cutoff $\Lambda$ of the Higgs sector is natural and stable under any radiative corrections from the UV physics of the Higgs sector, if the Higgs sector is embedded into the scale-invariant theory that breaks the scale symmetry spontaneously, leading to very light dilaton at low energy.

\subsection{Scale anomaly and the dilation effective potential}
\label{a1}
In the theory, where the scale symmetry is spontaneously broken, very light dilaton of mass $m_D^2\sim \left|{\cal E_{\rm vac}}\right|/f^2\ll \left|{\cal E_{\rm vac}}\right|^{1/2}$, arises as a Nambu-Goldstone boson, provided that the scale anomaly is much smaller than the scale of spontaneous breaking of the scale symmetry
\begin{equation}
-\left<\partial_{\mu}{\cal D}^{\mu}\right>=-\left<\theta_{\mu}^{\mu}\right>=-4{\cal E_{\rm vac}}\ll f^4\,.
\end{equation}
where ${\cal D}^{\mu}$ is the dilatation current and the dilaton decay constant $f$ is of the order of the spontaneous scale-symmetry breaking scale. Then, at low energy $E<M$, taking $\left|{\cal E_{\rm vac}}\right|\sim M^4$, one can write down the low-energy effective theory of dilaton that saturates the scale anomaly:
\begin{equation}
{\cal L}_D^{\rm eff}=\frac12\partial_{\mu}\chi\partial^{\mu}\chi-V_A(\chi)\,,
\end{equation}
where $\chi$ describes the small fluctuations around the asymetric vacuum,
\begin{equation}
\theta_{\mu}^{\mu}\approx 4{\cal E_{\rm vac}}\left(\frac{\chi}{f}\right)^4,
\label{dilaton}
\end{equation}
with $\left<\chi\right>=f$ at the vacuum. 

The dilatation current in the dilaton effective theory  is given as 
\begin{equation}
{\cal D}^{\mu}=\frac{\partial {\cal L}_D^{\rm eff}}{\partial(\partial_{\mu}\chi)}\left(x^{\nu}\partial_{\nu}\chi+\chi\right)-x^{\mu}{\cal L}_D^{\rm eff}\,.
\end{equation}
The scale anomaly then takes~\cite{Schechter:1980ak}, using the equations of motion for $\chi$,  
\begin{equation}
\partial_{\mu}{\cal D}^{\mu}=4V_A-\chi\frac{\partial V_A}{\partial\chi}\,.
\label{anomaly_equation}
\end{equation} 
From eqs.\,(\ref{dilaton}) and (\ref{anomaly_equation}) we get 
\begin{equation}
V_A(\chi)=\left|{\cal E}_{\rm vac}\right|\left(\frac{\chi}{f}\right)^4\left[4\ln\left(\frac{\chi}{f}\right)-c_0\right]\,.
\end{equation}
We note that the anomaly equation (\ref{anomaly_equation}) does not fix the constant $c_0$. But, 
our choice of the vacuum, $\left<\chi\right>=f$, fixes $c_0=1$. For the nonlinear realization of the dilatation symmetry we rewrite $\chi=fe^{\sigma/f}$ to get
\begin{equation}
{\cal L}_D^{\rm eff}=\frac12e^{2\sigma/f}\partial_{\mu}\sigma\partial^{\mu}\sigma-V_A(\sigma)\,,
\label{dilaton_lagrangian}
\end{equation}
with $V_A(\sigma)=\left|{\cal E}_{\rm vac}\right|e^{4\sigma/f}\left(4\sigma/f-1\right)$.

\subsection{Dilaton and scale invariance of the Higgs sector}
\label{a2}
To solve the fine-tuning problem of Higgs mass, we embed the Higgs sector to a scale-invariant theory in UV. The UV theory is assumed to break the scale symmetry spontaneously, generating dynamically a condensate $\left<\theta^{\mu}_{\mu}\right>\sim M^4$. The scale $M$ defines the intrinsic scale of the UV theory such as the dynamical mass in eq.~(\ref{ir scale}) or the scale of phase transitions in~\cite{Tavares:2013dga}.   

Integrating out all the modes above the dynamical scale $M$ in the Higgs UV sector, 
the low energy effective theory of the Higgs fields is given as, turning off all the SM interactions except the Higgs self interactions and the dilaton coupling,  
\begin{equation}
{\cal L}_H={\cal L}_D^{\rm eff}+\frac12\partial_{\mu}\phi^{\dagger}\partial^{\mu}\phi-M_{\phi}^2e^{2\sigma/f}\phi^{\dagger}\phi-\lambda\left(\phi^{\dagger}\phi\right)^2+\cdots\,,
\label{effective Higgs}
\end{equation}  
where the ellipsis denotes the higher order terms of $\phi^{\dagger}\phi$, suppressed by powers of $M$. Note that we have included in the effective theory the the dilaton coupling to the Higgs fields, as they come from the same UV dynamics, shown in section \ref{dilaton-higgs}.                              
All the effects above the UV cutoff, taken to be $M$, of the Higgs sector  are approximated in the effective theory  by two relevant and marginal operators, namely the Higgs quadratic coupling, $M_{\phi}^2$, and the Higgs quartic coupling, $\lambda$. 
Since the scale symmetry that protects the Higgs quadratic coupling is spontaneously broken, generating a scale $M$,  it should be of the order of $M$ or $M_{\phi}^2=c_mM^2$, if one integrates out all the heavy modes of $E>M$. 

Now we argue that, because of the scale symmetry that is spontaneously broken at low enegy, the Higgs quadratic coupling $M_{\phi}$ is unphysical just like the phase of nucleon mass in the coupling of pions to nucleons is unphysical because of the spontaneously-broken chiral symmetry\footnote{The intrinsic scale $M$ is also unphysical in this sense. The physical quantities are such as the ratio $M/\Lambda_{\rm SB}$ in eq.~(\ref{ir scale}) and the intrinsic scale at the vacuum, $Me^{\left<\sigma\right>/f}$, or the Higgs quadratic coupling at the vacuum, $M^2_{\phi}e^{2\left<\sigma\right>/f}$. We often do not specify $\left<\sigma\right>$ to the physical scale, when there is no confusion, since we choose $\left<\sigma\right>=0$.}.  To see this, we integrate out the heavy modes further down to $M^{\prime}<M$.  Neglecting the logarithmic corrections to $c_m$, the Higgs quadratic coupling becomes
\begin{equation}
{\cal L}_m=-c_m{M^{\prime}}^2e^{2\sigma/f}\phi^{\dagger}\phi\,.
\end{equation}
This change of the quadratic term can be compensated, if we shift the dilaton field as
\begin{equation}
\sigma\to\sigma^{\prime}=\sigma+f\,\ln\left(\frac{M^{\prime}}{M}\right)\,.
\end{equation}
Under the scale transformation $M\to M^{\prime}$ the dilaton potential has to change  as 
\begin{equation}
V_A(\sigma)=\left|{\cal E}_{\rm vac}\right| e^{4\sigma/f}\left(\frac{4\sigma}{f}-1\right)\to
V^{\prime}_A(\sigma)=\left|{\cal E}^{\prime}_{\rm vac}\right| e^{4\sigma/f}\left(\frac{4\sigma}{f}-1\right)\,,
\end{equation}
where ${\cal E}^{\prime}_{\rm vac}={\cal E}_{\rm vac}\left(M^{\prime}/M\right)^4$. 
In terms of the shifted dilaton field, $\sigma^{\prime}=\sigma+\sigma_0$ with $\sigma_0=f\ln\left(M^{\prime}/M\right)$, the dilaton potential becomes 
\begin{equation}
V^{\prime}_A(\sigma)=V_A(\sigma^{\prime})=\left|{\cal E}_{\rm vac}\right| e^{4\sigma^{\prime}/f}\left(\frac{4\sigma^{\prime}}{f}-1\right)\,.
\end{equation}
This transformation of the dilaton potential can be easily seen if one notes 
under the scale transformation $M\to M^{\prime}$ the scale anomaly transforms as with $\chi^{\prime}=e^{\sigma^{\prime}/f}$ 
\begin{equation}
\theta^{\mu}_{\mu}={\cal E}_{\rm vac}\left(\frac{\chi}{f}\right)^4\to {\theta_{\mu}^{\mu}}^{\prime}={\cal E}^{\prime}_{\rm vac}\left(\frac{\chi}{f}\right)^4={\cal E}_{\rm vac}\left(\frac{\chi^{\prime}}{f}\right)^4\,,
\end{equation}
and the anomaly equation becomes
\begin{equation}
{\theta^{\mu}_{\mu}}^{\prime}=4V_A(\chi^{\prime})-\chi^{\prime}\frac{\partial}{\partial\chi^{\prime}}V_A(\chi^{\prime})\,,
\end{equation} 
while the vacuum energy of the ground state or $\left<\theta^{\mu}_{\mu}\right>$ is left invariant. 
We note also that the kinetic term in the effective dilaton Lagrangian (\ref{dilaton_lagrangian}) is kept properly normalized by the anomaly equation\,(\ref{anomaly_equation}). 

We see that because of the shift symmetry of the dilaton field the Higgs sector is scale-invariant up to the logarithmic violation through the constant $c_m$ and the quartic coupling $\lambda$. Hence, as long as the shift symmetry of the dilaton is good enough, the Higgs quadratic coupling  $M_{\phi}^2$ should be unphysical. This property is not spoiled under any radiative corrections from the UV physics  of the Higgs sector with spontaneously broken scale-symmetry, because one can always compensate the radiative corrections by shifting the dilaton field, as we have shown in this appendix\,\ref{a2}.
The constraint on the Higgs mass, studied in~\cite{Tavares:2013dga},  therefore does not apply to our model that has light dilaton from the spontaneously broken scale-symmetry, noted also in~\cite{Ghilencea:2016dsl}.

\subsection{The renormalization condition $m_{\phi}^2(\Lambda)=0$}
Now we turn on the SM interactions of the Higgs fields, which will break the scale symmetry that the Higgs-dilaton sector enjoys. From the effective potential (\ref{pot}) or the effective Lagrangian density (\ref{effective Higgs}) we calculate the one-particle irreducible (1PI) effective potential for the Higgs fields by integrating out all SM particles and possibly some new particles to get at one-loop, neglecting the higher order terms, 
\begin{equation}
V_{\rm eff}(\sigma,\phi)=V_A(\sigma)+\left(M_{\phi}^2e^{2\sigma/f}-c_1\Lambda^2\right)\phi^{\dagger}\phi+\frac{\beta}{8}\left(\phi^{\dagger}{\phi}\right)^2
\left[\ln\left(\frac{\phi^{\dagger}\phi}{v_{\rm ew}^2}\right)-c_2\right]+c_4\,,
\label{1pi_pot}
\end{equation}
where the loop momentum is cut off at $\Lambda\sim M$ and the effective potential is expanded in powers of $\Lambda$ with their coefficients $c_i$ and $\beta$ being functions of Higgs couplings to SM particles and also to new additional heavy particles that the UV sector of Higgs fields might have\footnote{If one applies strictly to our discussion Bardeen's original proposal for the naturalness problem~\cite{Bardeen:1995kv}, the only consistent quadratic terms allowed in the radiative corrections in (\ref{1pi_pot}) are ones due to heavy particles associated with the UV sector of the Higgs fields, but not the one from the regulator. Here, for simplicity, without any confusion the correction $c_1\Lambda^2$ stands collectively for all radiative corrections to the quadratic term that the effective theory receives.  }. Though the scale symmetry is explicitly broken by SM interactions, one can still impose the renormalization condition (\ref{quadratic term}) that the Higgs quadratic term in the 1PI effective potential at $\Lambda$ vanishes by redefining the dilaton field $\sigma\to\sigma^{\prime}=\sigma+{\bar\sigma_0}$ with a suitable choice of $\bar\sigma_0$: 
\begin{equation}
m_{\phi}^2(\Lambda)\equiv \left.\frac{\partial^2V_{\rm eff}}{\partial\phi^{\dagger}\partial\phi}\right|_{\phi=0=\sigma^{\prime}}=M_{\phi}^2e^{-2{\bar\sigma_0}/f}-c_1\Lambda^2=0\,.
\label{r_condition}
\end{equation}

The choice of the renormalization condition, eq.\,(\ref{r_condition}) or eq.\,(\ref{quadratic term}), is consistent with the scale symmetry that the Higgs-dilaton Lagrangian of eq.\,(\ref{effective Higgs}) enjoys and also with the fact that the Higgs quadratic term is protected above the intrinsic scale $M$ of the UV sector by the symmetry\footnote{One may argue that the renormalization condition (\ref{r_condition}) is not compatible with the UV theory of the Higgs sector that has some new massive particles or nonperturbative scale. But, we emphasize that what matters is whether or not one can maintain the renormalization condition at all orders in perturbation. }.  We emphasize that the choice of $\bar\sigma_0$ in eq.\,(\ref{r_condition}) is equivalent to the choice of the counter term in the Coleman-Weinberg potential, since the quadratic term $M_{\phi}^2e^{-2{\bar\sigma_0}/f}\phi^{\dagger}\phi$ represents the effects of the UV sector of Higgs fields. Therefore, if we fix the UV cutoff the Higgs sector to be $\Lambda$, the intrinsic scale of the UV theory at vacuum is determined by the condition with $\left<\sigma\right>=0$
\begin{equation}
c_mM^2=c_1\Lambda^2\,.
\label{ren_point}
\end{equation}
We note that the renormalization condition eq.\,(\ref{ren_point}) holds for any cutoff $\Lambda$ because the Higgs quadratic term in the effective potential eq.\,(\ref{1pi_pot}) is scale-covariant:
Under the scale transformation $\Lambda\to \Lambda^{\prime}$ the dilaton field transforms $\sigma\to\sigma^{\prime}=\sigma+f\ln\left(\Lambda^{\prime}/\Lambda\right)$ and the quadratic term becomes
\begin{equation}
\left(M_{\phi}^2e^{2\sigma^{\prime}/f}-c_1{\Lambda^{\prime}}^2\right)\phi^{\dagger}\phi=
\left(\frac{\Lambda^{\prime}}{\Lambda}\right)^2\left(M_{\phi}^2e^{2\sigma/f}-c_1\Lambda^2\right)\phi^{\dagger}\phi\,.
\end{equation}
This is equivalent to saying that the Callan-Symmanzik  equation for the 1PI two-point function of Higgs fields in Fourier transforms becomes
\begin{equation}
\left(p\cdot\frac{\partial}{\partial p}+f\frac{\partial}{\partial\sigma}-2\right)\Gamma^{(2)}(p;\sigma)=0\,.
\end{equation}

\end{document}